\documentclass[review]{elsarticle}

\makeatletter
\def\ps@pprintTitle{
	\let\@oddhead\@empty
	\let\@evenhead\@empty
	\def\@oddfoot{\reset@font\hfil\thepage\hfil}
	\let\@evenfoot\@oddfoot
}
\makeatother

 \biboptions{comma,sort&compress}

\usepackage{graphicx}
\usepackage{amsmath}
\usepackage{here}
\usepackage{cuted}
\usepackage{hyperref}
\usepackage{tabularx}

\usepackage[dvips]{epsfig}

\begin{document}
\markboth{Stephan Narison, Montpellier (FR)}{ }
\begin{frontmatter}

\title{Phenomenology of a double dilaton soft-wall model: Alpha strong from Ricci flow and pion Form Factors at intermediate-energy region\,\tnoteref{invit}}
\tnotetext[invit]{Talk given at QCD25 - 40th anniversary of the QCD-Montpellier Conference.}
\author[a]{H\'ector Cancio}
\ead{hcancio@ifae.es}
\author[a]{Pere Masjuan}
\address[a]{\textit{\small{Grup de F\'isica Te\`orica, Departament de F\'isica, Universitat Aut\`onoma de Barcelona, and Institut de F\'isica d’Altes Energies (IFAE), and The Barcelona Institute of Science and Technology (BIST), Campus UAB, 08193 Bellaterra (Barcelona), Spain}}}


\date{\today}
\begin{abstract}

Through a holographic model of QCD, we present a phenomenological approach to study the running of the strong coupling constant $\alpha_s$ in both non-perturbative and perturbative regimes. The renormalization of the metric tensor, driven by the Ricci Flow, and the breaking of conformal and chiral symmetries -thanks to introducing a double dilaton model and large-$N_c$ corrections- allow us to relate the existence of an infrared fixed point in the coupling constant with a smooth matching to pQCD well above 2 GeV. This is done through a model with two fit parameters and one matching point. The proposed dilaton model yields linear Regge trajectories and decay constants for scalar, vector, and tensor meson families similar to their experimental counterparts. We finally study neutral and charged pion form factors to show an application of the running coupling constant obtained.

\end{abstract}
\end{frontmatter}
\newpage
\section{Introduction}
AdS/CFT conjecture has been used extensively in the search for a possible holographic model dual to QCD in the non-perturbative regime. First introduced as an equivalence between strongly coupled $\mathcal{N}=4$ super Yang-Mills theory and supergravity \cite{Maldacena, Gubser, Witten}, it was possible to break conformal symmetry to obtain approximate large-$N_c$ phenomenological models of QCD at low energies. In particular Hard \cite{Erlich,DaRold} and Soft Wall \cite{Karch} models provide a way to compute masses, decay constants, and Regge trajectories of mesons of arbitrary spin.

In this work, we are interested in obtaining the running coupling constant in the strongly coupled regime and extending it to the well-known perturbative one. We can get a QCD-like running from dual models like Einstein-dilaton gravity \cite{Gursoy,Kiritsis}, or by considering Light-Front QCD embedded in a AdS background \cite{Brodsky}. In our case we will relate it to the running of the metric tensor on the other side of the duality. Intuitively, an abrupt change of $\alpha_s$ at $\Lambda_{QCD}$ could be related to an expansion of spacetime on the holographic side so that the running of the metric tensor is driven by the Ricci Flow (\cite{Friedan}, \cite{Hamilton} and \cite{Perelman}). The procedure provides for a family of dilaton fields which beyond returning an $\alpha_s(Q)$, return masses and decay constants of vector mesons.
To show an application of the running coupling constant we turn on the tensions existing in neutral and charged pion form factors between experimental data \cite{BaBar, Belle, CELLO, CLEO, NA7, JLAB1, JLAB2, Dally1, Dally2, Dally3, Bebek74, Bebek76, Bebek78} and asymptotic limit predictions from pQCD \cite{LepageBrodsky}.

This letter is organized as follows. In Section \ref{Sec.RicciFlow}, we present some generalities of the Ricci Flow and its relation to $\alpha_s$. In Section \ref{Sec.DDSW}, we propose a new holographic QCD model, and we explore its running coupling constant. A discussion of its prediction for masses is presented in Section \ref{Sec.MassesDecays}. In Section \ref{Sec.PionFF} we study an application of the previous running coupling constant to neutral and charged pion form factors, and we finalize the letter with concluding remarks in Section \ref{Sec.Conclusions}.

\section{Ricci Flow}\label{Sec.RicciFlow}
The objective is to present the theoretical framework used to construct the model and establish a precise novel relation between the running of the metric tensor and the strong coupling constant $\alpha_s$. Consider the AdS metric tensor:
\begin{equation}
\label{eqn:AdSBackground}
ds^2=\frac{R^2}{z^2}\left(dz^2+\eta_{\mu\nu}dx^{\mu}dx^{\nu}\right).
\end{equation}
$R$ is the AdS radius, which is related to 't Hooft's coupling $\lambda$ by $\frac{R^2}{\alpha'}=\sqrt{\lambda}$, where $\lambda=g_{\text{YM}}^2N_c$. We are in the large-$N_c$ limit in the quantum field theory at the boundary of AdS, so we assume $\lambda \gg 1$. $g_{\text{YM}}$ is the QCD strong coupling. Since $\alpha_s=g_{\text{YM}}^2/(4\pi)$, $\alpha_s=\frac{R^4}{4\pi N_c \alpha'^2}$.
If there is exact conformal symmetry in the boundary theory, then $\alpha_s$ is constant and the background remains a pure AdS spacetime. To break conformal symmetry, we can consider a deviation from pure AdS imposing $R=R(z)$, thus considering $\frac{R^2}{\alpha'}=\sqrt{\lambda}$ holds approximately, allowing $\alpha_s$ to depend on the holographic coordinate. Returning to Eq.(\ref{eqn:AdSBackground}), we can write it in the form $ds^2=\frac{R^2}{z^2}dz^2+g_{\mu\nu}dx^{\mu}dx^{\nu}$,
where $g_{\mu\nu}=(R^2/z^2)\eta_{\mu\nu}$ and $R^2\sim\sqrt{\alpha_s}$. We can relate a running of $\alpha_s$ with a running of $g_{\mu\nu}$ taking into account $\mu= z^{-1}$ and differentiating $g_{\mu\nu}(\mu)$, we obtain
$\mu \frac{\partial g_{\mu\nu}}{\partial \mu}=2g_{\mu\nu}(\mu)\Bigl(1+\frac{\mu}{4}\frac{d}{d\mu}\log (\alpha_s(\mu^2))\Bigr)$.
In this way, we can relate naturally the running of $\alpha_s(\mu^2)$ with the running of the metric tensor $g_{\mu\nu}(\mu)$. This result motivates the introduction of the running of a general metric tensor as:
\begin{equation}
\frac{\partial g_{\mu\nu}}{\partial t}=2g_{\mu\nu}-2R_{\mu\nu},
\end{equation}
where $t=\log{\mu}$. This formula is known as the Ricci Flow. Introduced in the context of string theory \cite{Friedan}, and in mathematics \cite{Hamilton,Perelman}, Ricci Flow has been a method to compute the renormalization of non-linear sigma models. We will treat the holographic coordinate as the evolution parameter of the flow (see \cite{Jackson}). If $\alpha_s$ does not run at all, solving the Ricci Flow returns a metric tensor corresponding to AdS background.

In presence of a dilaton $\phi(z)$, solving the flow with an appropiate ansatz for $R_{\mu\nu}$ (see \cite{CancioMasjuan} for details), we obtain a $z$-dependent coupling constant:
\begin{equation}
\label{eqn:RunningDilaton}
\alpha_s(z)=\frac{1}{4\pi N_c \alpha'^2}e^{-\frac{4}{5}\phi(z)}.
\end{equation}
The above procedure allows us to define a general QCD running coupling constant for models with a dilaton. In other words, each holographic model will induce a particular running of $\alpha_s$. In the next section, we propose a dilaton model that will induce a QCD running coupling constant with an infrared fixed point and with a matching to pQCD. 

\section{Double Dilaton Soft Wall model and its running}\label{Sec.DDSW}
Our objective is to explore what possible dilaton fields would return an $\alpha_s$ resembling the one in QCD. Understanding vector and axial-vector mesons see different geometries \cite{Hirn}, we introduce a background with two dilatons, one with a positive and one with a negative sign, all together in a single model. This opposite signs will be responsible of chiral symmetry breaking and will enable us to study properties of the running of $\alpha_s$ at the same time, in particular the existence of an infrared fixed point. In this way, think of a background defined in $AdS \times AdS'$. To obtain a five-dimensional model we consider the diagonal $\Delta$ of $AdS\times AdS'$, defined as the submanifold $x=x'$.

The action on the diagonal is given by $S^{\Delta}=\int d^5x\frac{e^{-\phi^{\Delta}(z)}}{z^5}\left(-\frac{1}{2g_5^2}\left(F_L^2+F_R^2\right)\right)$. This choice not only gives a five-dimensional holographic model but also breaks chiral symmetry: $AdS$ contains a $U(N_f)_L$ theory and $AdS'$ contains a $U(N_f)_R$ theory, being $N_f$ the number of flavours. The product $AdS\times AdS'$ contains a theory with gauge symmetry $U(N_f)_L\times U(N_f)_R$. By selecting the diagonal one obtains a $U(N_f)_V$ theory, breaking chiral symmetry geometrically. The \textit{diagonal} dilaton obtained is given by $\phi^{\Delta}(z)=\log\left(2\cosh(\lambda^2(z^2+k))\right)$.
The constant $\lambda$ is introduced to render the argument of $\cosh$ unitless, we have identified $k=k'$ and we have chosen $k=1 \text{ GeV}^{-2}$ for simplicity. From now on we will refer to this model as the Double Dilaton Soft Wall model (DDSW).

\begin{figure}
\begin{center}
\includegraphics[scale=0.55]{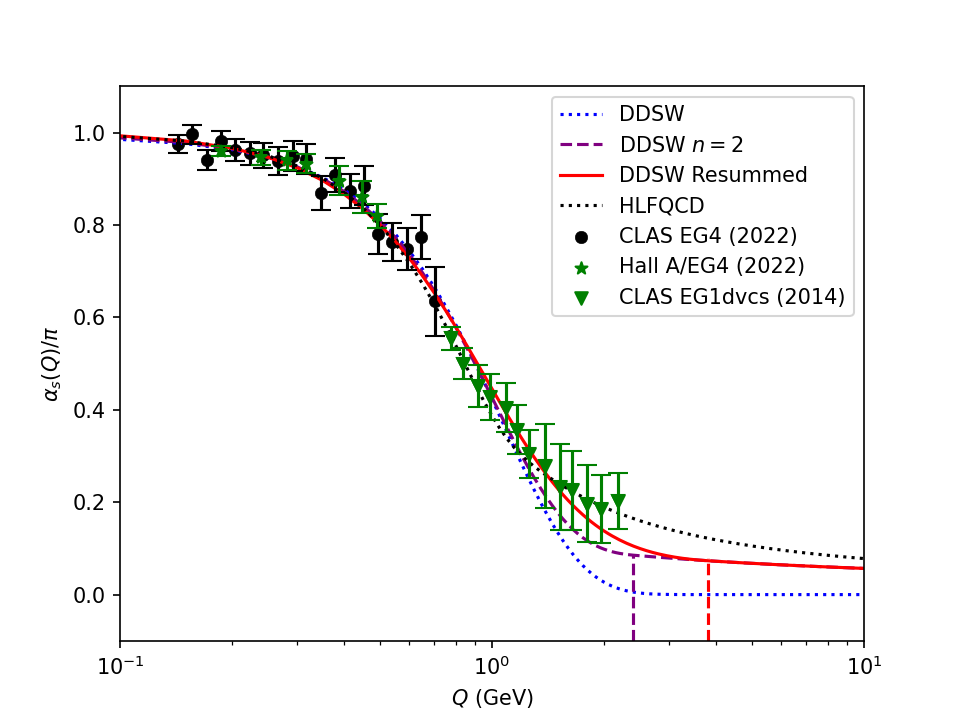}
\caption{Model from Eq.(\ref{eqn:parametrization}) (dotted blue), matching to perturbative QCD running coupling using DDSW model Eq.(\ref{eqn:largeN}) for $n=2$ (dashed purple), resummed (red) Eq.(\ref{eqn:Resum}). Their $\chi^2/DOF$ values are $1.45$, $0.99$ and $0.93$, respectively. We also compare with the holographic light-front model \cite{Brodsky} (dotted black). The dashed vertical lines indicate the matching point with perturbative QCD for $n=2$ and the resummed version (so for $Q>2.39$ and $Q>3.79$, respectively, we have perturbative QCD running coupling in the same color).}
\label{figure:running}
\end{center}
\end{figure}

In Ref.\cite{Brodsky} the running of $\alpha_s$ at low energies was obtained with the embedding of Light-Front QCD into a Soft Wall model with a positive dilaton, obtaining the so-called Holographic Light-Front QCD model. In our case, since no intrinsic scale is used in holographic models of QCD, the distinction between low and high energy is blurred, yet a matching with pQCD should be possible. We will explore such matching from the large-$N_c$ expansion. As usual we identify $\mu \sim Q$ so we can define $\alpha_s(Q)\sim\alpha_s(1/z)$. The DDSW model predicts:
\begin{equation}
\label{eqn:parametrization}
\frac{\alpha_s(Q)}{\pi}=\frac{a}{\cosh^{4/5}{(b(Q^2+1))}}.
\end{equation}
We have redefined the constants that normalize the function and render the $\cosh$ function unitless. The parameters $a$ and $b$ are determined by data fitting, using the effective strong coupling $\alpha_{g1}(Q)/\pi$ collected in Ref.\cite{Deur}, obtained from JLab experiments Hall CLAS EG4 (from Q = 0.143 GeV to 0.704 GeV), CLAS EG4/E977110 (from Q = 0.187 GeV to 0.490 GeV) and EG1dvcs (from $Q=0.775 \text{ GeV}$ to $2.177 \text{ GeV}$), represented in Fig.\ref{figure:running} as solid star, solid circle, and solid triangle, respectively. Notice the normalization at the fixed-point $\alpha_s(Q \simeq 0)/\pi \simeq 1$ in Fig.\ref{figure:running}. From a fit to data with the model in Eq.(\ref{eqn:parametrization}), we obtain $a = 1.545 \pm 0.047$ $b=1.150 \pm  0.041 \text{ GeV}^{-2}$ with a reasonable $\chi^2/DOF = 1.45$. As can be seen in Fig.\ref{figure:running}, we obtain naturally an infrared fixed point. Even though the fit is reasonably good, we do not find pQCD running coupling constant at high energies. This is expected since holographic models of QCD work only in the strongly coupled regime.

We have used a holographic model, so large-$N_c$ corrections to this result are expected, allowing a matching to pQCD. These large-$N_c$ corrections are difficult to derive from first principles. Following a more phenomenological approach, observe this expansion can be understood as quantum corrections to the AdS/CFT dictionary.
Working the above expression, we obtain:
\begin{equation}
\label{eqn:largeN}
\frac{\alpha_s(Q)}{\pi}=c\sum_{g=0}^{n-1} \left(\frac{a}{\cosh^{4/5}{(b(Q^2+1))}}\right)^g.
\end{equation}
Thanks to the $g=0$ term we can match to perturbative QCD running at four loops with reference scale $M_Z=91.18 \text{ GeV}$, $\alpha_s(M_z)=0.118$. The parameters $a,b,c$ must be determined either from data or from matching to pQCD or a combination of both. In the following, we opt for matching only the parameter $c$ and let the data tell us about $a$ and $b$ and normalize at $Q_0$ both image and derivative of the low and high energy results for $\alpha_s$. In this way we determine the $c$ parameter solving a system of equations, while a fit to the data will determine $a$ and $b$ parameters in an iterative procedure. We solve the system recursively, starting with the initial value $c=\alpha_s(M_z)/\pi = 0.118/\pi \simeq 0.0376$, for then fit the data to obtain $a$ and $b$. Solving the system we determine $Q_0$.

We can add more and more terms to the partial sum in Eq.(\ref{eqn:largeN}). Since it is a geometric series it can be fully resumed assuming the ratio is less than unity (suggested by the fitted coefficients), to obtain:
\begin{equation}
\label{eqn:Resum}
\frac{\alpha_s(Q)}{\pi}=\frac{c}{1-a \text{ sech}^{4/5}(b(Q^2+1))}.
\end{equation}
In this case, and with the fitted coefficients $a=0.962(1)$, $b=0.297(7) \text{ GeV}^{-2}$, the matching point is moved up to $Q=3.79 \text{ GeV}$ with $c=0.070$ and $\chi^2/DOF=0.93$, see Fig.\ref{figure:running}. Interestingly, the resummation has a slower decrease at around $1-2$ GeV, allowing a better fit to the data.

\section{Masses and decay constants}\label{Sec.MassesDecays}

\begin{figure}
\begin{center}
\includegraphics[scale=0.55]{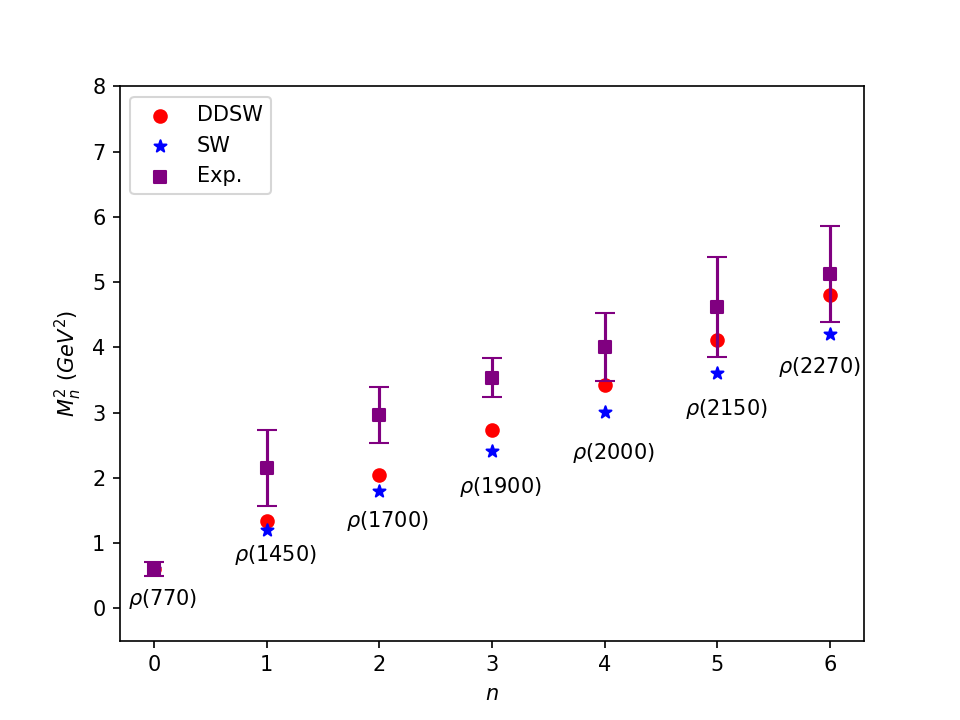}
\caption{Vector meson Regge trajectories $(M_n^2,n)$ for the DDSW model (red points), the SW model (blue stars), and comparison with experimental data (purple boxes) \cite{Workman}, \cite{Navas}.}
\label{figure:regge}
\end{center}
\end{figure}

The proposed dilaton can be used to predict different meson Regge trajectories. We study the trajectories for scalar, vector, and tensor mesons by solving its correspoding equations of motion (same form as in \cite{Karch} using our diagonal dilaton). To solve them, the lightest mass must be fixed to be the $\rho$-meson mass $M_0=M_{\rho}=775 \text{ MeV}$, which corresponds to $\lambda=0.4152$. Regge trajectories for our DDSW model are linear as in the Soft Wall model, that is, $M_n^2\sim n$.  
We depict our predictions using the DDSW model for the $\rho$-meson family (red points in Fig.\ref{figure:regge}) together with the physical values from the PDG \cite{Workman} (purple boxes) where errors are obtained via de half-width rule \cite{MasjuanArriolaBroniowski}, and the prediction of Soft Wall model \cite{Karch} (blue stars).
Comparison to experimental data demands merging all $\rho$ states with total angular momentum $J=1$, including 1S and 2S states, as the holographic model cannot distinguish among them. The Regge slope for the radial trajectory for the $\rho$ family read $0.87(7)$GeV$^{-2}$ from Ref.\cite{Masjuan:2012gc}. The DDSW model yields a slope $0.6962(18)$ GeV$^{-2}$.

\begin{figure}[htbp]
    \centering
    \begin{minipage}[t]{0.48\textwidth}
        \centering
        \includegraphics[width=\linewidth]{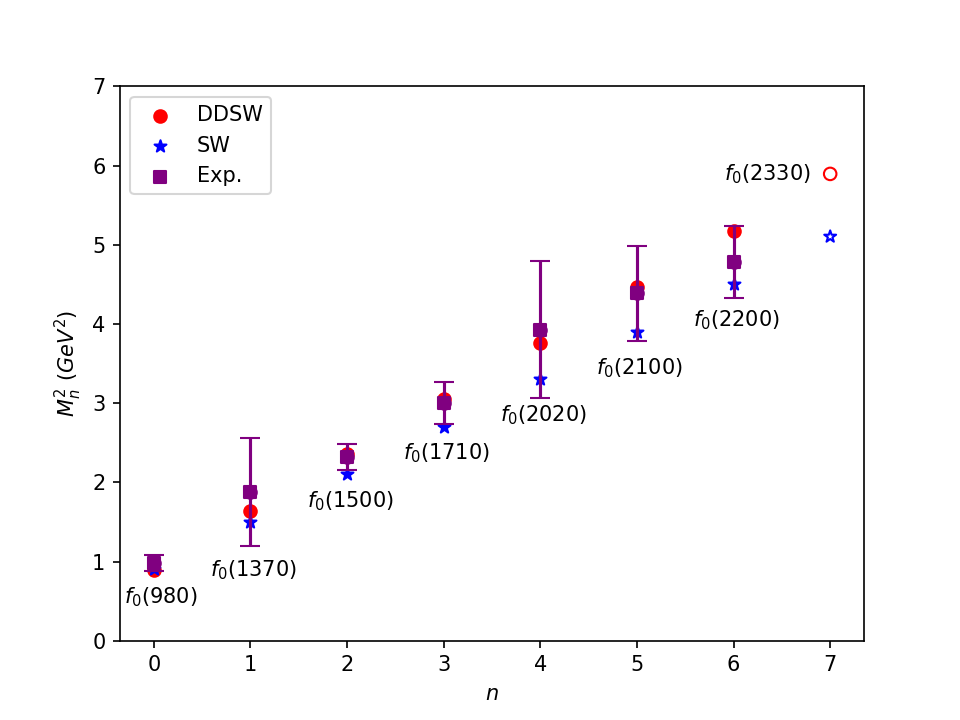} 
        \caption{Scalar meson Regge trajectories $(M_n^2,n)$ for the DDSW model (red points), the SW model (blue stars), and comparison with experimental data (purple boxes) \cite{Navas}.}
        \label{figure:reggeScalar}
    \end{minipage}
    \hfill
    \begin{minipage}[t]{0.48\textwidth}
        \centering
        \includegraphics[width=\linewidth]{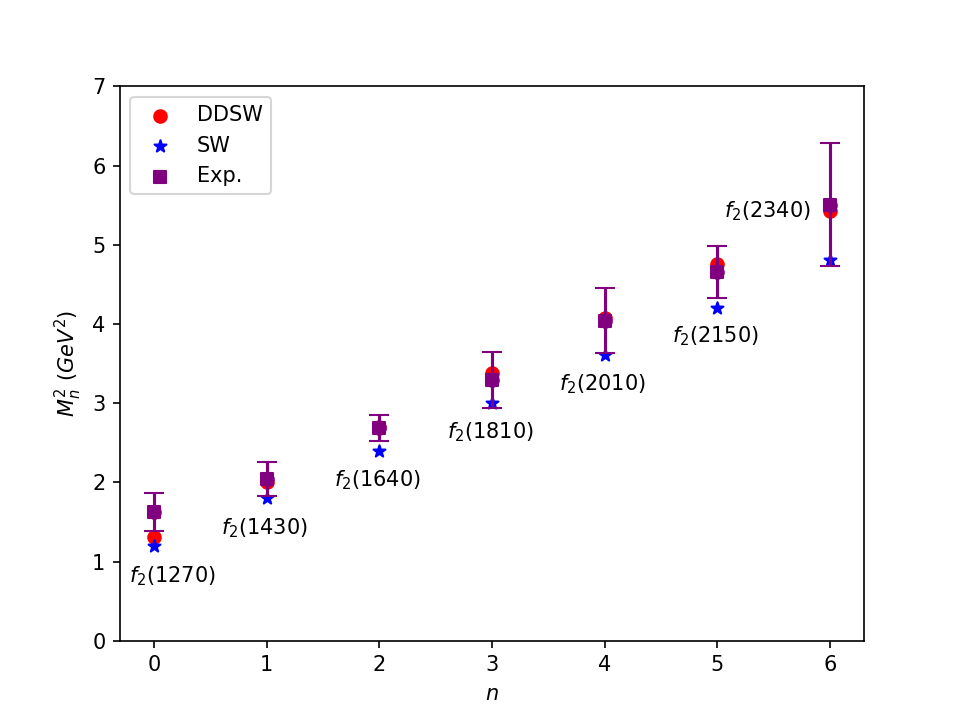} 
        \caption{Proposal for tensor meson Regge trajectory $(M_n^2,n)$ for the DDSW model (red points), the SW model (blue stars), and comparison with experimental data (purple boxes)\cite{Navas}.}
        \label{figure:reggeTensor}
    \end{minipage}
\end{figure}

Observe scalar resonances with $n=2, 3, 5$ show a departure to the experimental counterpart below $1\%$. In Figure \ref{figure:reggeScalar}, the empty circle and the empty star represent the prediction of DDSW and Soft Wall models, respectively, for $f_0(2330)$ meson. We have not included this resonance as a experimental data point since there is no consensus for its mean value \cite{Navas}. We predict the mass of the $f_0(2330)$ as $2428 \text{ MeV}$ in the DDSW model, while we observe this value is similar to that of \cite{Rodas}, which gives a mass of $2419\pm 64\text{ MeV}$, heavier than other analysis present in the PDG \cite{Navas}.

About the tensor resonances, some studies distinguish four different Regge trajectories \cite{MasjuanArriolaBroniowski}, \cite{Anisovich}. The output of our model is a single trajectory, so we have compared our predicted masses with the experimental data available. Our model is a large-$N_c$ model, so we have stable mesons, with zero decay widths. As consequence, we have grouped the resonances found in experiment if they are overlaped by its decay width. In this way, the Regge trajectory in Table \ref{figure:reggeTensor} is a proposal of organising the $f_2$ resonances.

The prediction of the $\rho$-meson decay constant using the DDSW model is also possible as is related to the residue at the pole $M_0$ on the equation of motion. It reads $F_{0}^{1/2}=F_{\rho}^{1/2}=346 \text{ MeV}$ which departs from the PDG value $F_{\rho}^{1/2}=348 \text{ MeV}$ by $0.57\%$ only \cite{Workman}. This is an improvement with respect to the Soft Wall (SW) result, which reads $F_{\rho}^{1/2}=260 \text{ MeV}$ \cite{Leutgeb}.
From the $\rho$ decay constant predicted by our DDSW we can obtain $\Gamma(\rho^0\to e^+e^-)=6.87 \text{ keV}$. This gives a departure of $2.41\%$ in comparison to the experimental value $7.04(6)\text{ keV}$ \cite{Workman}. For the usual Soft Wall model the value is $2.19 \text{ keV}$ with a departure of $68.9\%$.

\section{Neutral and charged pion form factors}\label{Sec.PionFF}
At this point let us discuss some applications of the running of the strong coupling constant studied in earlier sections into pion form factors \cite{CancioMasjuanWork}. In particular we study the neutral pion transition form factor as well as the charged pion form factor at spacelike momentum. In the last few years there have been a renewed interest in the intermediate-energy region of both form factors. In BaBar \cite{BaBar} and Belle \cite{Belle} experimental data points for the neutral pion transition form factor there is a tension with the pQCD limit \cite{LepageBrodsky}, while, the experimental data for the charged pion form factor also shows a deviation from its perturbative asymptotic decay. To shed light into this energy region, we employ the pion distribution amplitude formalism with the running coupling coming from the DDSW model.

Consider two distribution amplitudes $\phi_{\pi^0}$ and $\phi_{\pi^{\pm}}$ which define the neutral $F_0(Q^2)$ and charged $F_{\pm}(Q^2)$ pion form factors as: 
\begin{equation}
\label{eqn:neutralFFdef}
Q^2F_0(Q^2)=\frac{\sqrt{2}f_{\pi}}{3}\int_0^1 \frac{dx}{x}\phi_{\pi^0}(x,Q^2), \qquad
Q^2F_{\pm}(Q^2)=\frac{8\pi f_{\pi}^2 \alpha_s(Q^2)}{9}\left(\int_0^1\frac{dx}{x}\phi_{\pi^{\pm}}(x,Q^2)\right)^2.
\end{equation}
Here, $f_{\pi}=0.131 \text{ GeV}$ is the pion decay constant. In the case of the neutral pion distribution amplitude, we assume an expansion for high $Q^2$ in terms of Gegenbauer polynomials $C^{3/2}_{2n}(2x-1)$ inspired by the one derived in \cite{LepageBrodsky}: 
\begin{equation}
\label{eqn:NeutralPDAExpansion}
\phi_{\pi^0}(x,Q^2)=\phi_{\text{asym}}(x)\sum_{n=0}^{\infty} b_{n}(Q^2)\hat{\alpha}_s(Q^2)^{\gamma_n}C^{3/2}_{2n}(2x-1)
\end{equation}
where $x$ is the momentum fraction and $\phi_{\text{asym}}(x)=6x(1-x)$ is the asymptotic limit of the function $\phi_{\pi^0}(x,Q^2)$ when $Q^2\to\infty$. $b_n$ are the expansion coefficients $\gamma_n$ are anomalous dimensions, with $b_0=1$. $\hat{\alpha}_s$ is the non-perturbative strong running coupling worked out in earlier sections \ref{eqn:Resum}. The above expansion \ref{eqn:NeutralPDAExpansion} implies:
\begin{equation}
\label{eqn:neutralFFExpansion}
Q^2F_0(Q^2)=\frac{\sqrt{2}f_{\pi}}{3}\sum_{n=0}^{\infty}c_n\hat{\alpha}_s(Q^2)^{\gamma_n}
\end{equation}
In this way our parametrization is considered a prediction coming from our $\hat{\alpha}_s$ function at first order. The $c_n$ coefficents are related with the previous ones in the pion distribution amplitude as $b_n=c_n/3$. The constant $c_0$ is fixed at 3 in order to obey the asymptotic limit
$\lim_{Q^2\to\infty}Q^2F_0(Q^2)=\sqrt{2}f_{\pi}$  \cite{LepageBrodsky}.

For the charged pion form factor, consider the analogous expansion in the pion distribution amplitude \cite{LepageBrodsky}:
\begin{equation}
\phi_{\pi^\pm}(x,Q^2)=\phi_{\text{asym}}(x)\sum_{n=0}^{\infty} \tilde{b}_{n}(Q^2)\hat{\alpha}_s(Q^2)^{\gamma_n}C^{3/2}_{2n}(2x-1)
\end{equation}
Using the previous definition for the charged pion form factor, we obtain:
\begin{equation}
\label{eqn:chargedFFExpansion}
Q^2F_\pm(Q^2)=16\pi F_{\pi}^2\hat{\alpha_s}(Q^2)\left|\sum_{n=0}^{\infty}\tilde{c}_n\hat{\alpha}_s(Q^2)^{\gamma_n}\right|^2
\end{equation}
with the relation between coefficients given as $\tilde{b}_n=\tilde{c_n}$, with $F_{\pi}=f_{\pi}/\sqrt{2}$. Note, however, that the asymptotic limit of the charged pion form factor reads
$\lim_{Q^2\to\infty}F_{\pm}(Q^2)=16\pi F_{\pi}^2\alpha_s(Q^2)$ \cite{LepageBrodsky}, being $\alpha_s(Q^2)$ the perturbative QCD strong coupling constant. That is, the asymptotic limit does not fix the first coefficients of the proposed expansion. Therefore, we will need a fit to experimental data in order to obtain these coefficients. Since it is a high $Q^2$ expansion, we use only the experimental data from \cite{Bebek74}, \cite{Bebek76} and \cite{Bebek78}.

\begin{figure}[htbp]
    \centering
    \begin{minipage}[t]{0.48\textwidth}
        \centering
        \includegraphics[width=\linewidth]{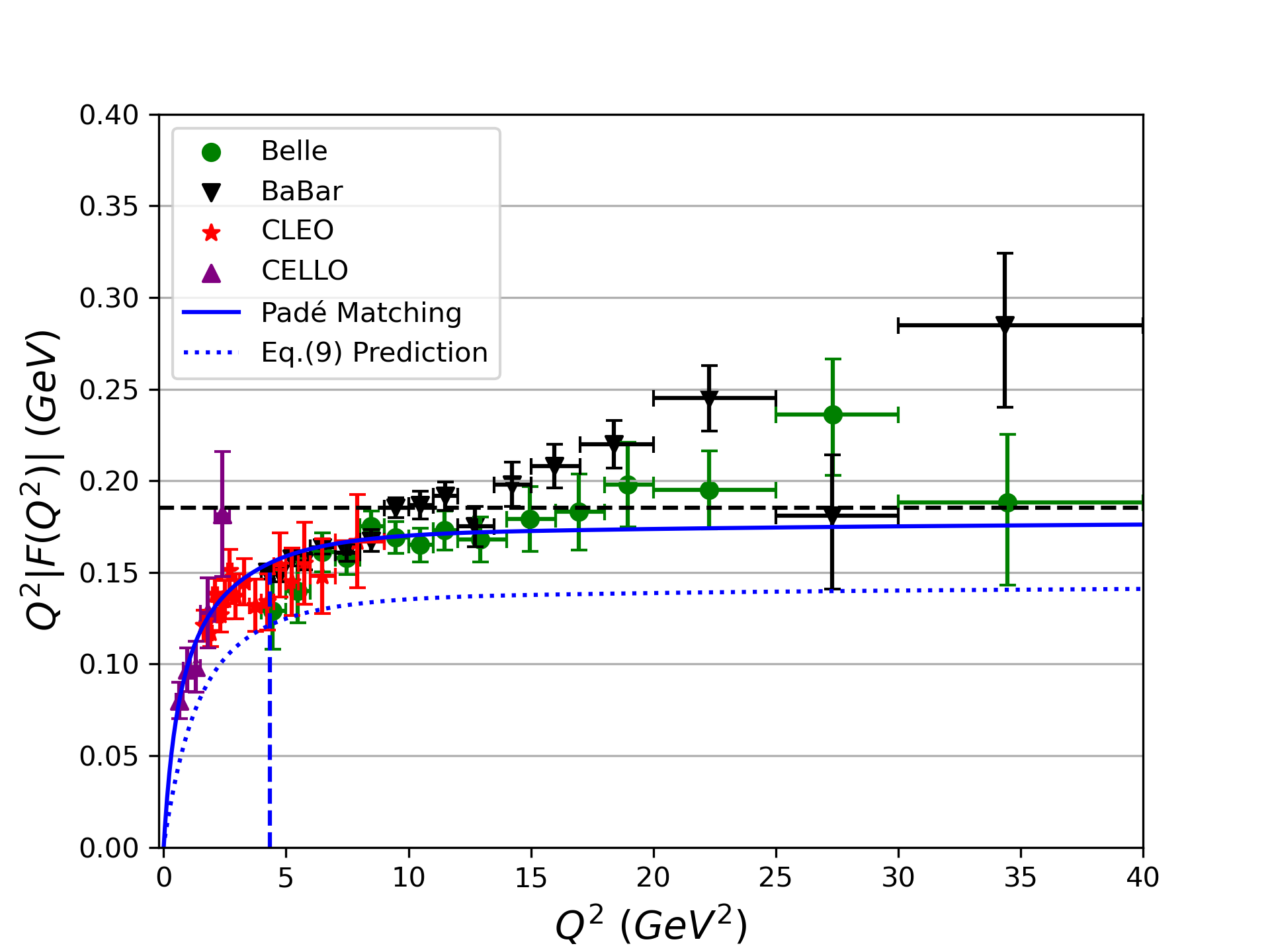} 
        \caption{Neutral pion electromagnetic form factor. Prediction from Eq.(\ref{eqn:neutralFFExpansion}) in dashed blue and result from matching procedure described in the text at low energies in solid blue. Fit to data using the same parametrization in solid blue. Experimental data are in green dots (Belle \cite{Belle}), black triangles (BaBar \cite{BaBar}), red stars (CLEO \cite{CLEO}) and purple triangles (CELLO \cite{CELLO}).}
        \label{fig:neutralFF}
    \end{minipage}
    \hfill
    \begin{minipage}[t]{0.48\textwidth}
        \centering
        \includegraphics[width=\linewidth]{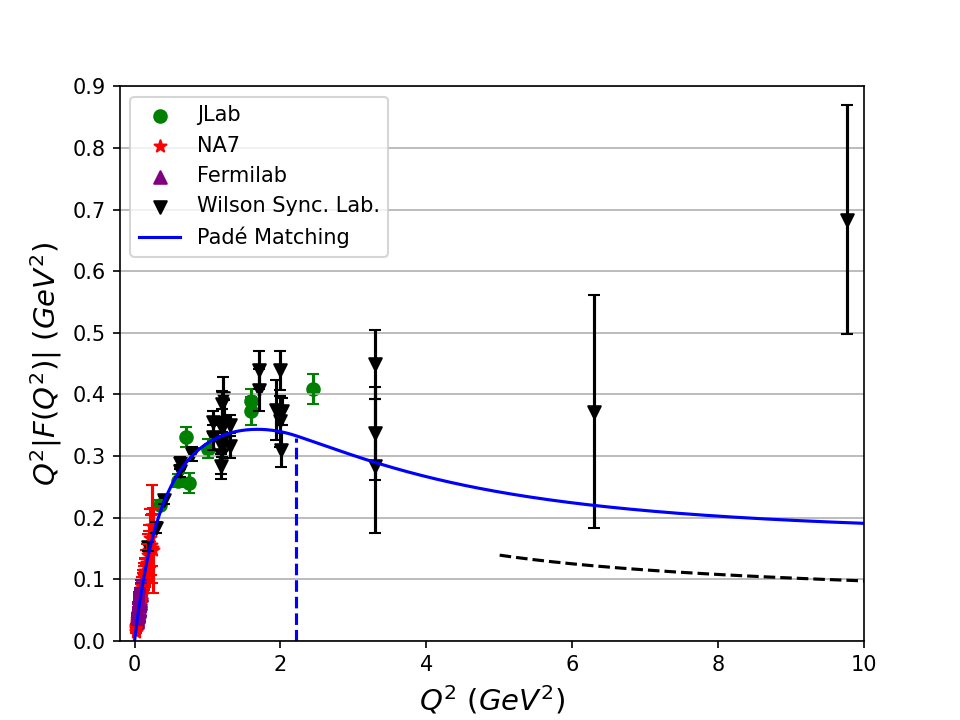} 
        \caption{Charged pion form factor. Solid blue curve is the result of using Eq.(\ref{eqn:chargedFFExpansion}) at high energies with a matching procedure described in the text at low energies. Experimental data in green dots (JLab \cite{JLAB1}, \cite{JLAB2}), red stars (NA7 \cite{NA7}), purple triangles (Fermilab \cite{Dally1}, \cite{Dally2}, \cite{Dally3}) and black triangles (Wilson Sync. Lab. \cite{Bebek74}, \cite{Bebek76} and \cite{Bebek78}).}
        \label{fig:chargedFF}
    \end{minipage}
\end{figure}

The prediction for the neutral pion transition form factor is seen in Fig.(\ref{fig:neutralFF}), corresponding to Eq.(\ref{eqn:neutralFFExpansion}) at $n=1$ with $c_0=3$ (from imposing the asymptotic limit) and $c_1=-c_0=-3$ (imposing the graph should pass through the origin). As can be seen in the figure, the expansion based in our running coupling constant allows us to extend the region of validity of the expansion to low energies. The experimental data is underestimated by imposing the limits $Q^2\to0$ and $Q^2\to \infty$. Therefore, we draw the solid blue curve treating Eq.(\ref{eqn:neutralFFExpansion}) with $n=1$ as a fit with parameters $c_0$ and $c_1$. By doing such a fit without imposing any additional condition we obtain $c_0=3.616\pm 0.083$ and $c_1=-3.2\pm0.20$. For the low energy region we use a [2, 1] Padé approximant and a matching procedure imposing continuity and differentiability at the matching point $Q_0^2$ with an unknown $a_4\neq 0$, obtaining $Q_0^2=4.35 \text{ GeV}^2$ and $a_4=1.4\cdot 10^{-3}$ from fitting the data \cite{CELLO}, \cite{CLEO}.

About the results in the charged pion form factor \ref{fig:chargedFF}, we consider a fit to the experimental data. We employ the experimental data from \cite{Bebek74}, \cite{Bebek76} and \cite{Bebek78}. We obtain $c_0=3.075\pm0.055$ and $c_1=-2.664\pm0.074$. To study the intermediate energy region, we employ a matching formalism \cite{NogueraVento}, so at low energies an adequate Padé approximant is used. We fit the experimental data \cite{NA7}, \cite{JLAB1} and \cite{JLAB2} with a $[4, 1]$ Padé approximant. As a result, we are able to obtain, after a matching procedure with the high energy result, a normalized $\chi^2$ function as close to $1$ as possible. We impose both functions must match at some unknown point $Q_0^2$ with an unknown $a_4\neq 0$. These two unknowns are fixed by imposing continuity and differentiability at $Q_0^2$ with an error of $10^{-11}$. By solving the resulting system of equations numerically we obtain $Q_0^2=2.21 \text{ GeV}^2$ and $a_4=-0.014$. As can be seen in the figure, the data is described by a curve with a maximum al low energies and then decreasing following the asymptotic limit for the charged pion form factor. In black discontinuous line the pQCD prediction is showed. Our fit remains above this pQCD prediction, a feature that was also obtained previously in other numerical approaches \cite{ChangCloet}.

\section{Conclusions and outlook}\label{Sec.Conclusions}
In this work, we have explored a new holographic model of QCD, named Double Dilaton Soft Wall model (DDSW), which from the Ricci flow allow us to obtain a parameterization for $\alpha_s$ able to match an infrared fixed point at low energies with pQCD at high energies. The matching with pQCD, which is found way above 2 GeV, is successful thanks to including large-$N_c$ corrections in the holographic side. In the process, we use experimental data from Ref. \cite{Deur} to fit the free leftover parameters of the $\alpha_s$ parameterization. Because our model Eq.(\ref{eqn:Resum}) have various adjustable parameters which limits its predictive power, we expect to improve it in the future.

As a by-product, we have studied the spectrum of vector mesons in the DDSW model by solving the equation of motion \cite{Karch} of different meson families with the dilaton mentioned above. This allowed us to determine meson masses, obtaining linear Regge trajectories for scalar, vector and tensor mesons. In particular, the predicted scalar and vector meson masses are heavier than in the usual Soft Wall model, thus closer to experimental results. Based on these results, we have done a proposal for the tensor meson Regge trajectory. Moreover, the squared root of the decay constant of the $\rho$-meson is improved compared to usual SW model, giving a departure of less than $1\%$ compared to the experimental result.

Finally we have seen applications of the running coupling constant previously studied in the neutral and charged pion form factors, in which tension with experimental data in the intermediate energy region do exist. In particular we have studied fits compatible with the asymptotic limits as well as the available data points.

These results open the possibility of exploring other phenomenological results. In particular, our method could be applied to other dilaton backgrounds, obtaining strong coupling runnings at low energies and corresponding matching to pQCD.

\section*{Acknowledgements} 
We would like to thank the conference organizers for helpful discussions. 
We gratefully acknowledge the Ministerio de Ciencia e Innovación for their financial support, which made the participation in the conference possible.
The work of P.M. was supported by the Ministerio de Ciencia e Innovación under grant PPID2023-146142NB-I00, by the Secretaria d’Universitats i Recerca del Departament d’Empresa i Coneixement de la Generalitat de Catalunya under grant 2021 SGR 00649, and by the Spanish Ministry of Science and Innovation (MICINN) through the State Research Agency under the Severo Ochoa Centres of Excellence Programme 2025–2029 (CEX2024-001442-S). IFAE is partially funded by the CERCA program of the Generalitat de Catalunya.

\nocite{*}
\bibliographystyle{unsrt}

\end{document}